\documentclass[prb,twocolumn]{revtex4-1}

\usepackage{amsmath}
\usepackage{amsfonts}
\usepackage{amssymb}
\usepackage{graphicx}

\begin{document}

\title{Casimir effect from a scattering approach}
\author{Gert-Ludwig Ingold}
\email{gert.ingold@physik.uni-augsburg.de}
\affiliation{Institut f{\"u}r Physik, Universit{\"a}t Augsburg,
     Universit{\"a}tsstra{\ss}e 1, D-86135 Augsburg, Germany}
\author{Astrid Lambrecht}
\email{astrid.lambrecht@lkb.ens.fr}
\affiliation{Laboratoire Kastler Brossel, CNRS, ENS, UPMC,
     Campus Jussieu Case 74, F-75252 Paris Cedex 05, France}
\date{\today}

\begin{abstract}
The Casimir force is a spectacular consequence of the existence of vacuum
fluctuations and thus deserves a place in courses on quantum theory. We argue
that the scattering approach within a one-dimensional field theory is well
suited to discuss the Casimir effect. It avoids in a transparent way
divergences appearing in the evaluation of the vacuum energy. Furthermore, the
scattering approach connects in a natural manner to the standard discussion of
one-dimensional scattering problems in a quantum theory course. Finally, it
allows to introduce students to the methods employed in the current research
literature to determine the Casimir force in real-world systems.
\end{abstract}

\maketitle

\section{Introduction} \label{sec:introduction}

The non-vanishing ground-state energy of the harmonic oscillator within a
quantum description represents a prominent example of the consequences of the
Heisenberg uncertainty relation.  However, this ground state energy is not
directly accessible. In the quantum field theory of the electromagnetic field,
where the field modes can be thought of as harmonic oscillators, the infinite
number of modes even leads to an infinite ground state energy.  This infinite
energy is usually removed by so-called normal ordering where in an operator
product creation operators $a^\dagger$ are moved to the left and annihilation
operators $a$ are moved to the right without accounting for the commutation
relation between $a^\dagger$ and $a$. On the level of a single harmonic
oscillator of frequency $\omega$, normal ordering amounts to replacing the
Hamiltonian
\begin{equation}
\label{eq:hHarmonicOscillator}
H = \hbar\omega\left(a^\dagger a+\frac{1}{2}\right) =
\frac{\hbar\omega}{2}\left(a^\dagger a+aa^\dagger\right)
\end{equation}
by
\begin{equation}
H = \hbar\omega a^\dagger a\,.
\end{equation}
The expectation value of the normal-ordered Hamiltonian in the ground state is
zero.

The situation changes when boundary conditions are imposed. For the
electromagnetic field one can imagine placing mirrors into space. Then, the
ground-state energy or vacuum energy, as it is usually called in the context of
a field theory, will take on a different albeit still infinite value. However,
one can ask how the vacuum energy changes when the boundaries are modified. A
well defined question is how this energy changes when two infinite parallel
mirrors are put into space. Changing their distance will lead to a change in
vacuum energy, or equivalently, the presence of the electromagnetic field even
in its ground state will result in a finite force on the two mirrors. In
1948, Casimir has shown that the force $F$ between two parallel ideal
mirrors at distance $L$ at zero temperature is given by \cite{Casimir1948}

\begin{equation} \label{eq:casimirForce3D} F_{3D} = \frac{\pi^2\hbar
c}{240}\frac{A}{L^4}\,.  \end{equation} Remarkably, apart from the distance
between the mirrors and their surface $A$, the Casimir force, as it is named
since then, depends only on physical constants, namely the Planck constant
$\hbar$ and the speed of light $c$.  A few years after the prediction by
Casimir first experiments aimed at its experimental verification
\cite{Derjaguin1951,Overbeek1951}. During the next three decades more Casimir
force measurements between two parallel plates or a plate and a spherical
surface followed, employing different methods and materials
\cite{Overbeek1954,Derjaguin1954,Derjaguin1956,Kitchener1957,Sparnaay1958,Black1960,Tabor1968,Arnold1972,Israelachvili1972,Blokland1978}.
Further insights into the historical context are given in
Ref.~\onlinecite{Milonni1992}. 

Modern measurement techniques such as precise torsion pendula, atomic force
microscopy or micro-electro-mechanical oscillators allow since more than
fifteen years for new precise Casimir force measurements.  Starting with the
experiments by Lamoreaux \cite{LamoreauxPRL1997} and Mohideen
\cite{MohideenPRL1998}   the Casimir effect has experienced an enormous
increase in experimental activities
\cite{EderthPRA2000,ChanPRL2001,ChenPRL2002,DeccaPRL2003,DeccaPRD2007,vanZwolAPL2008,ChanPRL2008,JourdanEPL2009,MasudaPRL2009,deManPRL2009,ChanPRL2010,TorricelliEPL2011,
SushkovNatPh2012,TangPRL2012,DeccaNatComm2013} as well as in theoretical
developments. Taking into account material properties, geometry, temperature
and the surface state in modern calculations is essential for obtaining
reliable theoretical predictions to be compared with Casimir force
measurements.  For a detailed discussion of recent developments, we refer the
reader to the collection of papers in Ref.~\onlinecite{DalvitLNP2011} and the
textbook by Bordag et al.\cite{Bordag2009} as well as to the resource letters
by Lamoreaux \cite{LamoreauxAJP1999} and Milton \cite{MiltonAJP2011} which can
serve as a guide to the literature.

The traditional way to determine the Casimir force (\ref{eq:casimirForce3D})
consists in calculating the ground state energy for all modes of the
electromagnetic field between two parallel ideal plane mirrors. In order to
handle the divergence of the vacuum energy, an appropriate high-frequency
cutoff procedure is employed.  Despite its formal character, this approach
following Casimir's original paper \cite{Casimir1948} is mostly taught and
described in textbooks, e.g. Ref.~\onlinecite{Itzykson85}. Of course there are
many other methods.  Those already presented in the American Journal of Physics
are based, e.g., on the calculation of the vacuum radiation pressure
\cite{HushwaterAJP1997} or on mode spectrum calculations for the force in one
dimension \cite{BoyerAJP2003}.

In research various complementary methods have been developed, such as  the
image method \cite{BrownPR1969} which may also be used to evaluate van der
Waals forces \cite{FarinaAJP2013}, vacuum radiation pressure calculations
\cite{JaekelJP1991} or, more recently, the worldline approach
\cite{GiesJHEP2003}, while multiple scattering techniques have been a valuable
tool in the context of Casimir physics since many years. For example Balian and
Duplantier used the technique to derive the Casimir energy for a sphere
\cite{BalianAnnPh1978}. The general formula to calculate the Casimir effect
that we will derive later on implies the logarithm of the determinant of the
scattering matrix. Such formulas appear already in early calculations of van
der Waals and Casimir interactions \cite{RennePhysica1971,LangbeinSTMP1974}.
Their appearance might be traced back to the Lippmann-Schwinger formulation of
scattering theory \cite{LippmannSchwingerPR1950}. More recently they have been
rediscovered using quantum optical scattering methods
\cite{JaekelJP1991,LambrechtNJP2006}, the $T$-operator approach
\cite{KennethPRA2006}, the Krein formula \cite{WirzbaPRD2006} and fluctuating
current scattering theory \cite{EmigPRL2007}, see also the short review by
Milton in Ref.~\onlinecite{MiltonJPA2008}.

Here, we want to propose the use of the scattering theory applied to a
one-dimensional field theory \cite{JaekelJP1991} as an alternative to the
traditional way of teaching the Casimir effect. Reducing the dimension of the
problem presents the advantage of avoiding the necessity to analyze all
electromagnetic modes between the two mirrors which unnecessarily complicates
the problem.  Of course, in one dimension the distance dependence will differ
from the $A/L^{4}$ behavior of the Casimir force (\ref{eq:casimirForce3D}) in
three dimensions. A simple dimensional argument allows us to deduce the
distance dependence of the force if only one space dimension is considered. As
one cannot define a surface in one dimension, the surface drops out of the
numerator and the Casimir force must scale as $1/L^{2}$. In the prefactor,
$\hbar c$ will be retained for dimensional reasons, while the numerical
prefactor will turn out to be different.  

It should be realized that the three-dimensional infinitely large plate-plate
configuration underlying the expression (\ref{eq:casimirForce3D}) for the
Casimir force refers to a very particular geometry never realized in
experiments.  In fact, in modern experiments the Casimir force is usually
measured between a sphere and a plate, thereby avoiding misalignment
\cite{Derjaguin1956}. A notable exception is the experiment described in
Ref.~\onlinecite{BressiPRL2002} where the force was measured between two finite
parallel plates.

Deriving the Casimir effect within a scattering theory as we will do in the
following, presents several pedagogical advantages. First of all, students who
have taken a first course in quantum mechanics are acquainted with
one-dimensional scattering problems. The scattering at a potential barrier, to
name but one example, is a standard exercise. While the formal scattering
approach is often not taught, it represents a natural extension of such
standard problems. The techniques acquired in this context can be useful in
other areas of modern physics like in mesoscopic physics \cite{Datta2003} where
the Landauer-B{\"u}ttiker theory \cite{BuettikerPRL1986} of the conductance constitutes
one example.

Secondly, the formal high-frequency regularization mentioned above is avoided.
Already Casimir had remarked in this respect: ``The physical meaning is
obvious: for very short waves (X-rays e.g.) our plate is hardly an obstacle at
all and therefore the zero point energy of these waves will not be influenced
by the position of this plate.'' \cite{Casimir1948}. A method taking into account
the physics at high frequencies is certainly preferable.  Furthermore, the
scattering approach allows to identify the contribution to the vacuum energy
depending on the distance between the two mirrors in a natural way. It thereby
clarifies the meaning of the Casimir energy.

Thirdly, the scattering approach has proven to be of great value in the
theoretical treatment of the Casimir effect. It allows to deal e.g. with
real mirrors described by a dielectric function and non-planar geometries.
The calculation presented here therefore gives an insight into methods
used in present-day research on the Casimir effect.

With this motivation for a scattering approach to the Casimir effect in mind,
some basic aspects of scattering theory needed in the sequel will be reviewed
in the following section. In Sec.~\ref{sec:CasimirScattering} we apply this
theory to obtain the change of the vacuum energy due to the presence of
scatterers in one dimension. For two scatterers the vacuum energy shift can be
decomposed into terms due to the individual scatterers and a term depending on
the distance of the two scatterers. The latter term is the Casimir energy which
is determined in Sec.~\ref{sec:twoScatterers}. The distance dependence of the
Casimir energy implies a force which we derive in Sec.~\ref{sec:casimirForce}
for the one-dimensional case. We conclude in Sec.~\ref{sec:outlook} by
sketching how the approach can be generalized to three dimensions and
geometries of practical interest. In Sec.~\ref{sec:problems} we have added
three problems which might be instructive for students.

\section{Scattering Theory}
\label{scatteringTheory}

\begin{figure}
\begin{center}
\includegraphics[width=0.8\columnwidth]{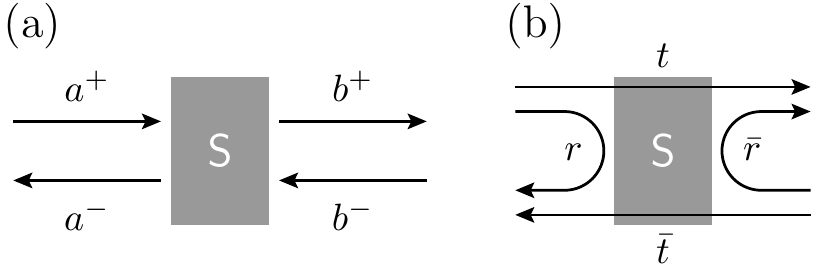}
\end{center}
\caption{Notation for the one-dimensional scattering problem where the gray area
indicates a finite scattering region. (a) $a$ and $b$ refer respectively to the
regions left and right of the scatterer. The superscripts $+$ and $-$ indicate
right- and left-going fields, respectively. (b) Reflection and transmission
amplitudes are denoted by $r$and $t$ for fields coming from the left and by
$\overline{r}$ and $\overline{t}$ for fields coming from the right.}
\label{fig:scatteringmatrix}
\end{figure}

We first review the basic properties of the scattering theory in one spatial
dimension. To this end, we assume that there exists a scattering region of
finite extent depicted in Fig.~\ref{fig:scatteringmatrix}(a) by the gray area.
In the regions to the left and to the right of the scattering region, plane
waves provide an appropriate solution.  Their amplitudes are $a^{\pm}$ and
$b^{\pm}$ with the index $\pm$ indicating right- and left-going waves, and $a$
and $b$ indicating the region to the left and to the right, respectively.

The scattering matrix $\mathsf{S}$ relates the ingoing waves characterized by
the amplitudes $a^+$ and $b^-$ to the outgoing waves with amplitudes $a^-$ and
$b^+$ according to
\begin{equation}
\label{eq:scatteringmatrix}
\begin{pmatrix} b^+\\ a^-\end{pmatrix} = \mathsf{S}
\begin{pmatrix} a^+\\b^-\end{pmatrix}\,.
\end{equation}
The scattering matrix as well as the amplitudes will in general depend on the
wave number $k$. For sake of simplicity, we will not make this dependence
explicit in most equations.

In the context of the Casimir effect, the scattering matrix will describe a
mirror with the diagonal and non-diagonal matrix elements referring to
transmission and reflection  processes, respectively. Accordingly, we choose
the following notation for the scattering matrix
\begin{equation}
\label{eq:Srt}
\mathsf{S} = \begin{pmatrix}t & \overline{r}\\r & \overline{t}\end{pmatrix}\,,
\end{equation}
where the scattering amplitudes visualized in
Fig.~\ref{fig:scatteringmatrix}(b) are not necessarily the same on both sides
of the scatterer.  This distinction leaves the possibility open that the mirror
behaves differently for modes arriving from the left or the right. It should be
kept in mind that $r,\overline{r}$ and $t,\overline{t}$ are reflection and
transmission amplitudes, respectively, and therefore in general complex
numbers. To avoid confusion, we note that here we follow the convention used in
quantum field theory which differs from the one commonly employed in mesoscopic
physics. There, the vector on the left-hand side of
(\ref{eq:scatteringmatrix}) is chosen as $(a^-, b^+)$ which has as a
consequence that the transmission amplitudes appear as off-diagonal elements
and not on the diagonal as in (\ref{eq:Srt}).

Current conservation requires the scattering matrix to be unitary
\begin{equation}
\label{eq:unitarity}
\mathsf{S}^\dagger\mathsf{S} = \mathsf{1}\,.
\end{equation}
As a consequence, the matrix elements have to obey the following three conditions
\begin{align}
\vert t\vert^2+\vert r\vert^2 =
\vert \overline{t}\vert^2+\vert \overline{r}\vert^2 &= 1
\label{eq:sMatrixProperty1}\\
r\overline{t}^*+t\overline{r}^* &= 0\,,
\label{eq:sMatrixProperty2}
\end{align}
where the star indicates complex conjugation. For later use we note that because
of the unitarity relations, the determinant of the scattering matrix can
be expressed solely in terms of the reflection coefficients
\begin{equation}
\label{eq:detSr}
\det(\mathsf{S}) = -\frac{r}{\overline{r}^*} = -\frac{\overline{r}}{r^*}\,.
\end{equation}

A simple example of a scattering matrix is given by
\begin{equation}
\label{eq:deltaS}
S = \begin{pmatrix}
1+r & r\\r & 1+r
\end{pmatrix}
\end{equation}
with the complex reflection coefficient
\begin{equation}
\label{eq:deltar}
r = \frac{g}{2ik-g}\,.
\end{equation}
With increasing wave number $k$, the scatterer turns from perfectly reflecting
into almost perfectly transmitting. The first two problems given in
Sec.~\ref{sec:problems} demonstrate how this specific scattering matrix can be
obtained in the contexts of single-particle quantum mechanics and of
electromagnetic transmission lines.

\begin{figure}
\begin{center}
\includegraphics[width=0.8\columnwidth]{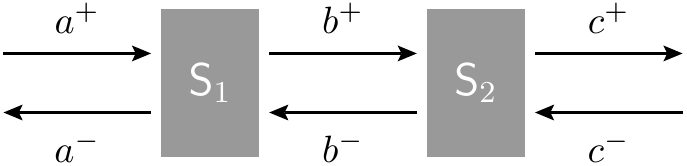}
\end{center}
\caption{Two scatterers with scattering matrices $\mathsf{S}_1$ and
$\mathsf{S}_2$ are used to describe a setup containing two mirrors with
no free field propagation in between.}
\label{fig:twoscatterers}
\end{figure}

In order to determine the Casimir force in a one-dimensional field theory,  we
need to describe a system composed of two scatterers separated by a distance
$L$. To approach this problem, we  first consider a set of two scatterers in
series as shown in Fig.~\ref{fig:twoscatterers} with no propagation in between.
It is convenient to consider the transfer matrix $\mathsf{T}$ instead of the
scattering matrix which relates the waves on the left-hand side to those on the
right-hand side of the scatterer according to
\begin{equation}
\label{eq:transfermatrix}
\begin{pmatrix} b^+\\ b^-\end{pmatrix} = \mathsf{T}
\begin{pmatrix} a^+\\ a^-\end{pmatrix}\,.
\end{equation}
In this way, the combined scattering properties of two or more scatterers in
series can easily be calculated.

As both, (\ref{eq:scatteringmatrix}) and (\ref{eq:transfermatrix}), represent
linear equations for the coefficients $a^{\pm}$ and $b^{\pm}$, it is
straightforward to convert the scattering matrix into the corresponding
transfer matrix and vice versa. One finds
\begin{equation}
\label{eq:mInTermsOfS}
\mathsf{T} = \frac{1}{\mathsf{S}_{22}}
\begin{pmatrix}
\det(\mathsf{S}) & \mathsf{S}_{12}\\
-\mathsf{S}_{21} & 1
\end{pmatrix}
\end{equation}
and
\begin{equation}
\label{eq:sInTermsOfM}
\mathsf{S} = \frac{1}{\mathsf{T}_{22}}
\begin{pmatrix}
\det(\mathsf{T})&\mathsf{T}_{12}\\
-\mathsf{T}_{21} & 1
\end{pmatrix}\,.
\end{equation}

If $\mathsf{S}_1$ and $\mathsf{S}_2$ are given, we can now employ
(\ref{eq:mInTermsOfS}) to obtain the corresponding transfer matrices
$\mathsf{T}_1$ and $\mathsf{T}_2$.  The transfer matrices allow us to express
the effect of the scatterers in series by means of a single transfer matrix
$\mathsf{T} = \mathsf{T}_2\mathsf{T}_1$ from which we obtain the effective
scattering matrix $\mathsf{S}$ for the two scatterers seen from the outside.
Its matrix elements expressed in terms of the reflection and transmission
amplitudes of the single scatterers read
\begin{equation}
\label{eq:smatrix2}
\mathsf{S} = \begin{pmatrix}
\displaystyle \frac{t_1t_2}{1-\overline{r}_1r_2} &
\displaystyle \overline{r}_2+\frac{\overline{r}_1 t_2 \overline{t}_2}
{1-\overline{r}_1r_2} \\[2em]
\displaystyle r_1+\frac{r_2 t_1 \overline{t}_1}{1-\overline{r}_1r_2} &
\displaystyle \frac{\overline{t}_1 \overline{t}_2}{1-\overline{r}_1r_2} 
\end{pmatrix}\,.
\end{equation}
Expanding the denominators in terms of a geometric series, one can convince
oneself, that all possible scattering processes including an arbitrary number
of back-and-forth scatterings between the two scatterers are contained in this
scattering matrix.

\section{Influence of scatterers on the vacuum energy}
\label{sec:CasimirScattering}

\begin{figure}
\begin{center}
\includegraphics[width=0.75\columnwidth]{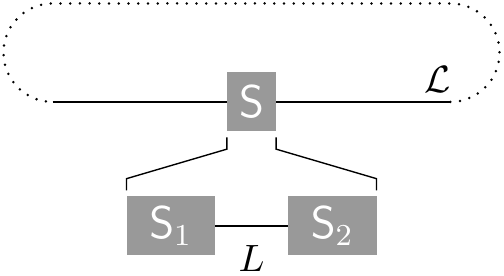}
\end{center}
\caption{A scatterer placed in a one-dimensional space of length $\mathcal{L}$
is discussed in Sec.~\ref{sec:CasimirScattering}. The dotted line indicates the
use of periodic boundary conditions. In Sec.~\ref{sec:twoScatterers}, the scatterer
will be replaced by two scatterers separated by a distance $L$ as indicated in the
lower part of the figure.}
\label{fig:scattering_pbc}
\end{figure}

So far, we have not addressed the question why the Casimir effect can be
approached by means of scattering matrices at all. Therefore, we now consider
the change of the vacuum energy when a scatterer is inserted into our
one-dimensional field. Later on, we will think of this single scatterer in
terms of two scatterers representing the two mirrors as indicated in the lower
part of Fig.~\ref{fig:scattering_pbc}. However, we know already that the two
scatterers can be described by a single scattering matrix and thus be viewed
from the outside as a single effective scatterer. Hence, for the moment it is
sufficient to consider one single scattering matrix $\mathsf{S}$.

We imagine the scatterer sitting in the middle of a one-dimensional space of
length $\mathcal{L}$ whose two ends are joined by periodic boundary conditions
(cf. upper half of Fig.~\ref{fig:scattering_pbc}). Ultimately, we will take the
limit $\mathcal{L}\to\infty$. The scatterer is characterized by a scattering matrix
$\mathsf{S}$ or, equivalently, a transfer matrix $\mathsf{T}$. The periodic
boundary condition can then be expressed as
\begin{equation}
\label{eq:eigenvalueCondition1}
\mathsf{T}\mathsf{T}_\mathcal{L}\begin{pmatrix}a^+\\a^-\end{pmatrix}
\overset{!}{=} \begin{pmatrix}a^+\\a^-\end{pmatrix}
\end{equation}
where
\begin{equation}
\label{eq:transferMatrixFreePropagation}
\mathsf{T}_\mathcal{L}=\begin{pmatrix}e^{ik\mathcal{L}} & 0\\
0 & e^{-ik\mathcal{L}}
\end{pmatrix}
\end{equation}
is the transfer matrix describing free propagation of a wave with positive wave
number $k$ moving to the right in the first and to the left in the second component.
From (\ref{eq:eigenvalueCondition1}) we obtain the eigenvalue condition for the
waves in the presence of a scatterer
\begin{equation}
\label{eq:eigenvalueCondition2}
\exp(-ik^\pm\mathcal{L}) = \frac{t+\overline{t}}{2}\pm
\left(\frac{(t+\overline{t})^2}{4}-\det(\mathsf{S})\right)^{1/2}
\end{equation}
where the scattering matrix is a function of $k$.

In view of the linear dispersion relation
\begin{equation}
\label{eq:dispersionrelation}
\omega=ck
\end{equation}
for the field modes of frequency $\omega$ and wave number $k$, the vacuum energy
is given by
\begin{equation}
E_\mathrm{vac} = \sum_n\frac{\hbar c}{2}(k^+_{(n)}+k^-_{(n)})
\end{equation}
where the sum runs over all modes. From (\ref{eq:eigenvalueCondition2})
we find
\begin{equation}
\label{eq:newKValues}
k^+ + k^-=2\frac{2\pi n}{\mathcal{L}}+\Delta k^++\Delta k^-\,.
\end{equation}
The first term refers to the case without scatterers and corresponds to the sum
of the wave numbers $2\pi n/\mathcal{L}$ of a pair of left- or right-moving
modes. The second and third terms are due to the presence of the scatterer
and according to (\ref{eq:eigenvalueCondition2}) given by
\begin{equation}
\Delta k^++\Delta k^-=i\frac{1}{\mathcal{L}}\ln[\det(\mathsf{S})]\,.
\end{equation}
Note that the determinant of the scattering matrix according to the unitarity
condition (\ref{eq:unitarity}) has modulus one. Therefore the logarithm is
purely imaginary, and the shift of the wave numbers turns out to be real as it
should.

The scatterer thus induces a shift in the vacuum energy of
\begin{equation}
\label{eq:changeInEnergy}
\begin{aligned}
\Delta E_\mathrm{vac} &= \frac{\hbar c}{2}\sum_n(\Delta k^++\Delta k^-)\\
&= \frac{\hbar c}{2}\int_0^\infty dk\frac{\mathcal{L}}{2\pi}
    \frac{i\ln[\det(\mathsf{S})]}{\mathcal{L}}\\
&= \frac{i\hbar c}{4\pi}\int_0^\infty dk\ln[\det(\mathsf{S})]\,.
\end{aligned}
\end{equation}
In going to the second line, we have made use of the fact that according to
(\ref{eq:newKValues}) the difference in wave number between subsequent
unperturbed modes amounts to $2\pi/\mathcal{L}$.

\section{Casimir energy for two scatterers}
\label{sec:twoScatterers}

If we place two scatterers into the one-dimensional field as depicted in
Fig.~\ref{fig:scattering_pbc}, we can make use of (\ref{eq:changeInEnergy}) to
obtain the change in the vacuum energy. We only need to determine the
determinant of the scattering matrix describing the two scatterers in a
distance $L$ from each other. The total transfer matrix
\begin{equation}
\label{eq:totalTransferMatrix}
\mathsf{T} = \mathsf{T}_L^{-1}\mathsf{T}_2\mathsf{T}_L\mathsf{T}_1 \,,
\end{equation}
which should be read from right to left, then contains four contributions: The
transfer matrix $\mathsf{T}_1$ of the first scatterer, the transfer matrix
for free propagation according to (\ref{eq:transferMatrixFreePropagation}) with
$\mathcal{L}$ replaced by $L$, the transfer matrix $\mathsf{T}_2$ of the
second scatterer, and finally the inverse of the transfer matrix
$\mathsf{T}_L$. The role of the last term becomes clear by inserting the
transfer matrix (\ref{eq:totalTransferMatrix}) into the eigenvalue condition
(\ref{eq:eigenvalueCondition1}) and realizing that
$\mathsf{T}_L^{-1}\mathsf{T}_\mathcal{L}=\mathsf{T}_{\mathcal{L}-L}$. The
original system of length $\mathcal{L}$ now consists of two parts of length
$\mathcal{L}-L$ and of length $L$. The last transfer matrix in
(\ref{eq:totalTransferMatrix}) thus ensures that the overall length of the
system remains constant even though a scattering region of length $L$ has been
inserted.

Performing the matrix multiplications, we find
\begin{equation}
\det(\mathsf{S}) = \frac{\det(\mathsf{S}_1)\det(\mathsf{S}_2)
-r_1 \overline{r}_2\exp(-2ikL)}{1-\overline{r}_1 r_2\exp(2ikL)}\,.
\end{equation}
The relations (\ref{eq:detSr}) now allow us to factorize the determinant
of the scattering matrix
\begin{equation}
\label{eq:detS}
\det(\mathsf{S}) = \det(\mathsf{S}_2)\det(\mathsf{S}_1)
\frac{1-\left[\overline{r}_1 r_2\exp(2ikL)\right]^*}{1-\overline{r}_1 r_2\exp(2ikL)}\,.
\end{equation}
This decomposition is physically quite significant because it implies that
the change of the vacuum energy (\ref{eq:changeInEnergy}) due to placing
mirrors into the field consists of three parts
\begin{equation}
\label{eq:deltaEVac}
\Delta E_\mathrm{vac} = \Delta E_\mathrm{vac}^{(1)}+\Delta E_\mathrm{vac}^{(2)}
+\Delta E_\mathrm{vac}(L)\,.
\end{equation}
The first two contributions to the determinant (\ref{eq:detS}) of the
scattering matrix and to the vacuum energy (\ref{eq:deltaEVac}) arise due to a
single mirror and they are therefore independent of $L$.  In contrast, the
third contribution depends on the distance $L$ because the coherent field
between the two mirrors is sensitive to their distance. Therefore, the third
term yields the Casimir energy. The last factor in (\ref{eq:detS}) is a pure
phase factor, so that the Casimir energy can be expressed as
\begin{equation}
\label{eq:deltaEVacL}
\Delta E_\mathrm{vac}(L) = \frac{\hbar c}{2\pi}\text{Im}\int_0^\infty dk\ln\left(
1-\overline{r}_1 r_2\exp(2ikL)\right)\,.
\end{equation}
The argument of the logarithm has a simple interpretation in terms of one
round-trip between the two scatterers consisting of a reflection at scatterer
1, a propagation over a distance $L$, a reflection at scatterer 2 and finally
another propagation over a distance $L$ in order to return back to scatterer 1.
Recalling the notation introduced in Fig.~\ref{fig:scatteringmatrix}(b), one
immediately sees that the force depends only on the inner reflection
coefficients $\overline{r}_1$ and $r_2$ of the two-scatterer set-up. 

\section{Casimir force in one dimension}
\label{sec:casimirForce}

The Casimir force is obtained from the distance-dependent part
(\ref{eq:deltaEVacL}) of the change of the vacuum energy induced by the two
scatterers as the derivative with respect to their distance $L$
\begin{equation}
\label{eq:casimirForceIntegral}
\begin{aligned}
F &= -\frac{d\Delta E_\mathrm{vac}(L)}{dL}\\
  &= \frac{\hbar c}{\pi}\mathrm{Re}\left[\int_0^\infty dk\,k
       \frac{\overline{r}_1 r_2\exp(2ikL)}{1-\overline{r}_1 r_2\exp(2ikL)}
       \right]\,.
\end{aligned}
\end{equation}

In order to evaluate the Casimir force it is convenient to rotate the axis of
integration from the positive real to the positive imaginary axis.  This step
is common in the evaluation of the Casimir force also for more general cases
\cite{GenetPRA2003,LifshitzJETP1956,DzyaloshinskiiAdvPhys1961}. Excluding the
special case of amplifying media, causality ensures that the integrand has no
poles in the upper complex half plane. This can be explicitly verified by
considering the scattering matrix (\ref{eq:deltaS}) with the reflection
coefficient (\ref{eq:deltar}). Furthermore, because of the exponential function
in the numerator, the integrand vanishes at infinity in the upper complex half
plane. Therefore, we can apply the residue theorem to turn the integration
contour. Substituting the wave number $k$ by $ix/2L$ one obtains
\begin{equation}
\label{eq:casimirForce1d}
F = -\frac{\hbar c}{4\pi L^2}\int_0^\infty dx\,x
      \frac{\overline{r}_1(x)r_2(x)\exp(-x)}{1-\overline{r}_1(x)r_2(x)\exp(-x)}
\end{equation}

Noting that the exponential in the numerator of (\ref{eq:casimirForce1d}) cuts
off the integrand, we may obtain the limit of perfect reflectors
$r_1=\overline{r}_2=-1$. The integral can be evaluated by first expressing the
integrand in terms of a geometric series and performing a resummation after
having integrated each term. We thus arrive at the Casimir force for perfect
reflectors in one dimension
\begin{equation}
F_\text{1D} = -\frac{\hbar c\pi}{24 L^2}\,.
\end{equation}
The sign implies an attractive force between the scatterers. The force scales
indeed as $1/L^{2}$, as expected from the dimensional argument presented in the
Introduction, with the same dependence on fundamental constants $\hbar$ and $c$
but a different numerical prefactor. This result has been obtained for
one-dimensional models in various contexts \cite{JohnsAPPB1975,BloetePRL1986}.

\section{Outlook}
\label{sec:outlook}

The calculations presented here may be generalized to three-dimensional space
involving the full electromagnetic field enclosed between two plane parallel
mirrors.  Then, the scattering on the mirror may still be described by a
$2\times 2$ scattering matrix relating the two outgoing fields to the two
incoming ones. However, now the electromagnetic field is characterized by its
frequency, transverse wave vector and polarization. From the symmetry with
respect to time translations and transverse space translations it follows that
these quantities are preserved throughout the scattering process. Therefore,
the expression (\ref{eq:changeInEnergy}) for the shift in the vacuum energy
still holds, provided an integration over the transverse wave vector and a
summation over the two polarizations is added. In this way, the scattering
approach leads to the original result (\ref{eq:casimirForce3D}) by Casimir.

In an arbitrary static configuration with two scatterers in vacuum, the Casimir
energy can still be written in the form (\ref{eq:changeInEnergy}). This
includes, e.g., the experimentally relevant cases of a sphere in front of a
plate
\cite{LamoreauxPRL1997,MohideenPRL1998,ChanPRL2001,DeccaPRL2003,vanZwolAPL2008,JourdanEPL2009,MasudaPRL2009,
deManPRL2009,TorricelliEPL2011,SushkovNatPh2012,TangPRL2012} and structured
surfaces \cite{ChenPRL2002,ChanPRL2008,ChanPRL2010,DeccaNatComm2013}. However,
in general, plane waves will no longer be adapted to the geometry of the
problem. The scattering processes now may lead to changes of the transverse
wave vector and to a coupling between polarizations resulting in
high-dimensional scattering matrices of a complex structure.

Apart from the geometry, the comparison with experimental results requires to
account also for material properties
\cite{LifshitzJETP1956,DzyaloshinskiiAdvPhys1961,LambrechtEPJD2000} and finite
temperature $T$
\cite{LifshitzJETP1956,DzyaloshinskiiAdvPhys1961,Sauer1962,MehraPh1967,SerneliusPRL2000}.
The former enter into the elements of the scattering matrix \cite{GenetPRA2003}
while the latter can be accounted for by replacing the vacuum energy
$\hbar\omega/2$ by the Planck factor \cite{Milonni1992}
$(\hbar\omega/2)\coth(\hbar\omega/k_BT)$.

Therefore, a calculation taking all these aspects into account can become
rather challenging and numerically demanding.

\section{Suggested Problems}
\label{sec:problems}

\textit{Problem 1.} Derive the scattering matrix (\ref{eq:deltaS}) with the reflection
coefficient (\ref{eq:deltar}) for a delta-like scattering potential 
\begin{equation}
V_0(x) = \frac{\hbar^2g}{2m}\delta(x)\,.
\end{equation}
Depending on the level of knowledge of quantum mechanics, different approaches
can be chosen, e.g.: (1) Start out from the textbook expression for the complex
reflection coefficient of a rectangular potential barrier of height $V_0$ and
width $a$. Take the limit $V_0\rightarrow\infty, a\rightarrow 0$ while keeping
$V_0a=\hbar^2g/2m$ constant to obtain (\ref{eq:deltar}). Derive the remaining
matrix elements of the scattering matrix by exploiting the symmetry of the
problem and the unitarity of the scattering matrix. Compare your result with
(\ref{eq:deltaS}).  (2) Start out from the boundary conditions at the
delta-like potential to obtain relations between the coefficients of the wave
functions on both sides of the scatterer (cf.  Fig.~\ref{fig:scatteringmatrix}).
Rearrange the equations so that you can read off the scattering matrix.

\textit{Problem 2.} Consider an infinite $LC$ transmission line with inductance $\bar L$
and conductance $\bar C$ per unit length. At $x=0$, the two conductors of the transmission
line are connected by an inductance $L$. Determine the reflection coefficient and prove
that it can be brought into the form (\ref{eq:deltar}). The symmetry of the problem
and the requirement of unitarity allow to arrive at the full scattering matrix
(\ref{eq:deltaS}).

\textit{Problem 3.} Convince yourself that the scattering matrix
(\ref{eq:smatrix2}) accounts for all possible scattering processes involving
two scatterers. Hint: Make use of a geometric series.

\begin{acknowledgments}
The authors thank Serge Reynaud for many insightful discussions. This work has
been supported by the DAAD and {\'E}gide through the PROCOPE program as well as
by the European Science Foundation (ESF) within the activity ``New Trends and
Applications of the Casimir Effect'' (www.casimir-network.com).
\end{acknowledgments}

\end{document}